\documentclass{aa}
\usepackage{graphicx}

\begin{document}
\title{Absence of linear polarization in H$\alpha$ emission of solar flares}
\author{M. Bianda\inst{1,2} \and  A.O. Benz\inst{2} \and
  J.O. Stenflo\inst{2,3} \and G. K\"uveler \inst{4} \and  R. Ramelli \inst{1} }

\offprints{mbianda@irsol.ch}

\institute{IRSOL, CH-6605 Locarno-Monti, Switzerland \and 
Institute of Astronomy, ETH-Zentrum, CH-8092 Z\"urich, Switzerland \and
Faculty of Mathematics and Science, University of Zurich, 8057
Z\"urich, Switzerland \and Fachhochschule Wiesbaden, Am Br\"uckweg 26,
D-65428 R\"usselsheim, Germany
}

\date{Received  July 13, 2004 ; accepted January 17, 2005 }
\markboth{Absence of linear polarization }{}

\abstract{High sensitivity observations of H$\alpha$ polarization of
  30 flares of different sizes and disk positions are reported. Both
  filter and spectrographic techniques have been used. The ZIMPOL
  system eliminates spurious polarizations due to seeing and
  flat-field effects. 
We didn't find any clear linear polarization signature above our
sensitivity level which was usually better than 0.1\%.
The observations include an X17.1 flare with 
gamma-ray lines reported by the RHESSI satellite. These results cast 
serious doubts on previous claims of linear polarization at the one percent 
level and more, attributed to impact polarization. The absence of linear
polarization limits the anisotropy of energetic protons in the
H$\alpha$ emitting region. The likely causes are isotropization by
collisions with neutrals in the chromosphere and defocusing by the
converging magnetic field.

}
\maketitle

\section{Introduction}
Linear polarization of the H$\alpha$ line emission of solar flares has been
reported in numerous papers of the past. H\'enoux \& Chambe (1990) and
H\'enoux et al. (1990) made pioneering contributions. Metcalf et
al. (1992, 1994) and Vogt \& H\'enoux (1999) found flare polarization with
imaging spectrographs, and in the latter work a correlation in time with soft
X-ray emission was reported. Vogt, Sahal-Br\'echot, \& H\'enoux (2002)
measured linear
polarization of 3--5\% in H$\alpha$ and H$\beta$ spectral observations of
flare kernels. 
Firstova et al. (1994) observed up to 20\% linear polarization at the Baikal
Solar Vacuum Telescope. 
Kashapova (2003) reported also observations of linear polarization measured
in moustaches or Ellerman bombs. Regular
observations of impact polarization are made  at the Ondrejov
Observatory (H\'enoux \& Karlick\'y 2003).
Hanaoka (2003) has recently recorded linear polarization with high
cadence and spatial resolution in a flare near disk center. He finds
values above the half percent level in most of the regions of
H$\alpha$ emission. The direction of linear polarization is
preferentially perpendicular to the flare ribbons, and the
polarization degree increases with H$\alpha$ brightness. Direction and
degree happen to be well correlated also with the spatial gradient of
the H$\alpha$ emission (Figs.~3 and 4 of Hanaoka 2003). As the total
H$\alpha$ flare emission correlates well with soft X-rays, the
correlation in time between the degree of polarization and H$\alpha$
brightness is consistent with the observations of Vogt \& H\'enoux (1999). 

Impact polarization by precipitating flare particles has been the
preferred interpretation since the early observations. As the observed
linear polarization appears to peak in the gradual flare phase and
significant polarization was reported at times outside impulsive hard
X-ray enhancements (H\'enoux et al. 1990), electrons were
excluded. Fletcher \& Brown (1995) exclude electrons also for theoretical 
reasons. Only flare accelerated ions may reach the level of H$\alpha $
emission in sufficient numbers. Thus beams of 
high-energy protons were proposed already by
the first observers. As protons with energy below 350 keV are
predicted to be most efficient in producing impact polarization (Vogt
\& H\'enoux 1996), its measurement would yield unique information on
protons accelerated in flares, a poorly known but possibly important
flare constituent. 

The measurement of linear polarization in the H$\alpha$ emission,
however, is extremely difficult due to the presence of gradients in space
and time. Spurious polarization signals
can originate from limited seeing, differential
optical aberrations and gain-table effects in
CCDs. Whenever the gradients are large, which occurs near peak flare
intensity, spurious polarization signals are the largest. 
Thus doubts about the reliability of the reported measurements
of linear polarization have been raised (Canfield \& Chang 1985; Fang
et al. 1995). 

A first set of imaging polarimetry measurements with higher sensitivity
performed in Locarno revealed no polarization signal from 
H$\alpha$ emission in flares above 
the few tenths of a percent level (Bianda et al. 2003).
The ZIMPOL (Zurich IMaging POLarimeter)
instrument was used at the IRSOL facilities. The main advantage of
ZIMPOL is that seeing and gain-table errors do not generate spurious
polarization signals, which allows the system to achieve unprecedented
polarimetric accuracy. Furthermore, the Gregory
Coud\'e telescope at IRSOL has low instrumental polarization that is
constant over the day. This combination allowed for a polarimetric accuracy and
time resolution substantially better than in the measurements reported
by other authors. In the first set of measurements performed
at IRSOL (Bianda et al. 2003) it was not possible to 
observe a major flare event, and the
H$\alpha$ filter allowed to study only the central part of the
spectral line and not the behavior in the wings.

In the present paper we present new observations, including
measurements taken with the spectrograph, in particular during 
the X17 flare on 28th October 2003.

\section{Instrument and observations}
ZIMPOL (Povel 1995; Gandorfer et al. 2004) 
allows to measure polarization with high
modulation rate (84 kHz for linear, 42\,kHz for circular), much higher
than the typical seeing frequencies (which are below about 1\,kHz). This
eliminates seeing-to-polarization cross talk, which is one of the
most important sources of spurious polarization signals in solar 
regions with high intensity gradients and rapidly evolving
structures. The polarimeter is a single-beam system, and the same CCD
pixels are used to measure all Stokes parameters.  
Instrumental spurious effects are thus minimized, and no flat-field technique
is needed to obtain accurate polarization images.
 
The 45 cm aperture Gregory Coud\'e telescope in Locarno has a
200\arcsec\ field of view. It is well suited for polarimetric observations, 
since instrumental polarization, mainly due to two folding
mirrors, is low and depends only on the Sun's declination. The effects
of the two folding mirrors mutually cancel around the equinoxes, so the
instrument is virtually polarization free  
(Wiehr 1971; S\'anchez Almeida et al. 1991).

The polarimetric observations were obtained with two techniques: 
imaging with a narrow bandpass filter, and spectrography with the
Czerny-Turner spectrograph.

In both cases, the ZIMPOL analyzer is the first optical device placed 
after the exit of the telescope along
the hour axis. It consists of a piezoelastic modulator and a
polarizer, which can be rotated in order to choose the orientation of the
linear polarization measurement.  The set-up used allows to record
simultaneously the intensity, one linear polarization component, and
the circular polarization $V/I$. A Dove prism image rotator
compensates for image rotation.

\subsection{Two-dimensional spatial polarimetry}

For spatial polarimetry observations, a 0.6\AA\ FWHM H$\alpha$ filter
is placed in front of the Dove prism.  A beam splitter feeds two
instruments: the Fachhochschule Wiesbaden (FHSW) flare detector system
(K\"uveler et al. 
2003) and the ZIMPOL CCD camera. Telecentric optics demagnifies the
image on the sensors.

The most promising active area, where a flare is expected, is
monitored. Priority is given to regions close to the limb, where
the largest observable impact polarization is expected. 
The polarization optics is set in such a way that the linear 
polarization oriented parallel to the closest limb is measured 
as positive $Q/I$.
Negative signatures are thus expected (Vogt \& H\'enoux 1996).

The FHSW flare detector system stores intensity images (12 per
second) and reads the GPS signal to 
provide the precise time of the observation.
Data are continuously stored in a buffer of programmable length
(typically 5 minutes) as long as no events are detected.  
If the intensity exceeds a predefined threshold,
the recording is triggered.
Data in the buffer are then permanently stored
together with the recordings of the following time
interval of typically 15 minutes. The trigger signal is sent to ZIMPOL
together with the precise GPS time information. 
The ZIMPOL software has been adapted to communicate with the FHSW 
system and to allow the data storage with the same buffer technique.
ZIMPOL data are stored with a rate of typically one image 
every four seconds. Before or after a flare
observation, calibration data and dark frames are
recorded. 
In the resulting images one pixel subtends $1\arcsec \times 1\arcsec$.

\subsection{Spectrographic mode} 
In the spectrographic mode the active regions are surveyed with the 
slit jaw plane viewer and the H$\alpha$ image of the guiding telescope.
The spectrum is recorded by the ZIMPOL camera.
The optical components are rotated in order to have the slit parallel
to the closest limb of an active region and such that positive $Q/I$ is 
defined to be directed along the slit. When
flare is visually detected, the telescope is moved in order to record data
from the erupting area. The measurements of $Q/I$ and $U/I$ are
obtained separately, rotating the analyzer
by 45$^\circ$. In some cases the complete Stokes
vector, otherwise only $Q/I$ and $V/I$, were recorded. 

\subsection{Spurious signature sources} 
The known possible sources of spurious polarization have been examined.
The instrumental polarization, which consists of a polarization
offset and a cross talk between different Stokes components, can be easily
compensated for in the data reduction
process, since its variation over a day is negligible.
Parasitic light could affect dark frames and introduce 
intensity-to-polarization cross talk and other spurious signatures.
Some of our first observations were corrupted by this problem, which could be
solved with better baffling. From other kinds of ZIMPOL measurements
we know that very strong intensity gradients can introduce spurious
signatures of order a few tenths of percent in the polarization
images, and are seen only in a few of the pixels of the CCD.
Another possible source of error could be related to the adjustments
of the optical setup. Its correctness was verified in most observations by 
feeding a signal of known polarization into the system before the analyzer 
for a few seconds.

Some further tests were performed to check the ability of ZIMPOL
to detect signatures evolving fast in time. 
Artificially modulated polarization signals inserted in front of the analyzer 
were correctly measured.
Test recordings were made to obtain H$\alpha$ magnetograms of active regions by
recording $V/I$ with the filter positioned
in the blue and red wings as well as in the line core. The resulting circular
polarization patterns were consistent with SOHO/MDI magnetograms.

\section{Results}\label{sec:results}
The observed 30 flares are summarized in Table \ref{listevents}, where
we give the date, 
the GOES classification in soft X-rays, the time of maximum soft X-ray
flux, the $\mu=\cos\theta$ position on the disk 
($\mu$ noted as $<0.0$ means outside
the disk), 
the noise level and the highest observed polarization value, 
the last two entries will be described later in more details.
In contrast to the measurements reported by other authors, a clear impact
polarization signal was never found.

\begin{table}[h!]
\caption[]{List of flare events observed. GOES flare classification 
(soft X-ray peak flux), position on the disk, and linear H$\alpha$ 
polarization noise and peak value are given.}
\label{listevents}
\begin{center}
\begin{tabular}{crccrr}
\hline\hline\noalign{\vskip2pt}
\multicolumn{6}{c}{Filter data} \\
\noalign{\vskip2pt}\hline\noalign{\vskip2pt}
Date	    &   Classi- &UT & $\mu$ & Noise& Max \\
    &  fication &Max & & level [\%]& Pol.  [\%] \\ 
\hline\noalign{\vskip3pt}
04   Jul 2002   &  C2.4	&10:45&	0.23 	& 0.02 &  -  \\
04   Jul 2002   &  C3.0	&13:34&	0.34 	& 0.03 &  -  \\
04   Jul 2002   &  C7.1	&14:57& 0.35	& 0.02 &  -  \\
05   Jul 2002   &  M3.2	&13:26&	0.07 	& 0.05 &  -  \\
07   Jul 2002   &   SF	&13:03&	0.66 	& 0.03 & 0.05   \\
11   Jul 2002   &  C2.1	&07:13&	0.41 	& 0.03 &  -  \\
11   Jul 2002   &  C4.0	&11:21&	0.43 	& 0.03 & 0.06   \\
11   Jul 2002   &  C2.5	&12:08& 0.43	& 0.02 & 0.03   \\
11   Jul 2002   &  M1.0	&14:19&	0.47 	& 0.03 & 0.05   \\
11   Jul 2002   &  M5.8	&14:48&	0.52 	& 0.02 & 0.07   \\
18   Jul 2002   &  X1.8	&07:24&	0.81 	& 0.02 & -  \\
02   Aug 2002   &  M1.0	&10:51&	0.45 	& 0.03 & 0.04   \\
04   Aug 2002   &  C7.4	&07:14&  $<0.0$ & 0.07 & -  \\
04   Aug 2002   &  M6.6	&10.33&	$<0.0$  & 0.07 & -  \\
04   Nov  2003  &  C5.7 &11:19&  $<0.0$ & 0.04 & -  \\
04   Nov  2003  &  M1.1 &13:45& $<0.0$  & 0.04 & -  \\
05   Nov   2003 &  M5.3	&10:52& $<0.0$  & 0.11 & -  \\
18   Nov  2003  &  M3.9 &08:31& 0.96	& 0.02 & -  \\
19   Nov  2003  &  C8.8 &08:17& 1.0 	& 0.02 & 0.03  \\  
\noalign{\vskip2pt}\hline\hline \noalign{\vskip3pt}
\multicolumn{6}{c}{
Spectrograph observations}\\
\noalign{\vskip2pt}\hline \noalign{\vskip3pt}
Date    &   Classi- &UT & $\mu$ & Noise& Max \\
    &  fication &Max & &  level [\%]& 	Pol.  [\%]
\\ \hline\noalign{\vskip3pt}
11  Jun  2003   &    C8.3   &08:30& 0.49   & 0.09 & - \\      
11  Jun  2003   &    M1.1   &10:33& 0.48   & 0.07 & - \\
11  Jun  2003   &    M1.4   &11:09& 0.48   & 0.08 & - \\
11  Jun  2003   &    M2.7   &13:21& 0.48   & 0.10 & - \\
12  Jun  2003   &    C7.2   &08:12& 0.2    & 0.08 & - \\
12  Jun  2003   &    C8.4   &08:29& 0.87   & 0.09 & - \\
12  Jun  2003   &    C8.1   &10:27& $<0.0$ & 0.40 & - \\
12  Jun  2003   &    C7.1   &11:47& 0.2    & 0.08 & - \\
12  Jun  2003   &    M1.0   &14:03& $<0.0$ & 0.20 & - \\
27  Oct   2003  &    M6.7   &12:43& 0.86   & 0.05 & - \\
28  Oct   2003  &    X17.1  &11:10& 0.95   & 0.05 & - \\      
\noalign{\vskip2pt}\hline
\end{tabular}
\end{center}
\end{table}                 

\begin{figure}
\begin{center}
\leavevmode
\resizebox{8.5cm}{!}{\hspace{0cm}\vspace{0cm}\includegraphics{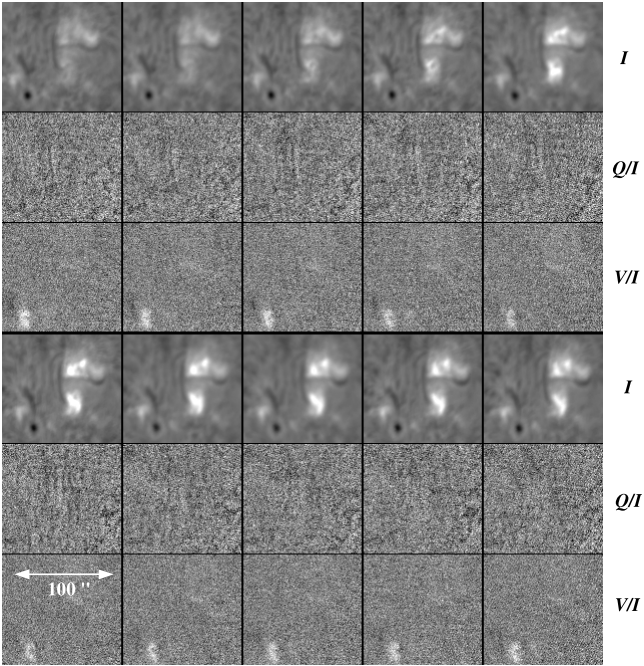}}
\end{center}
\vskip 0 cm
  \caption[]{Images of three Stokes parameters during a flare on 
7th July 2002. Time progresses from left to right in two minute steps
and continues in the lower half.}
  \label{fig:1}
\end{figure}

Figure \ref{fig:1} shows the evolution in time of the flare 
that occurred on 7th July 2002 at S19E46.
The recordings started at 12:50
UT and terminated at 13:11 UT. A total of 320 exposures were taken every 4
seconds, each of them with 2\,s integration time. Each square represents a
solar disk area of $120\arcsec \times 110\arcsec$. The first and
fourth row contain the H$\alpha$ intensity images, the second and
fifth row $Q/I$, while the third and sixth row show $V/I$. The $Q/I$
and $V/I$ images have been obtained by averaging 10 subsequent
exposures. The intensity image corresponds to a single exposure (in
the the middle of the time interval). The only faint polarization
signature that can be clearly detected is in $V/I$. It is due to the
Zeeman effect at the position of a sunspot (dark feature in the $I$ images). 

\begin{figure}
\begin{center}
\leavevmode
\resizebox{8.5cm}{!}{\hspace{0cm}\vspace{0cm}\includegraphics{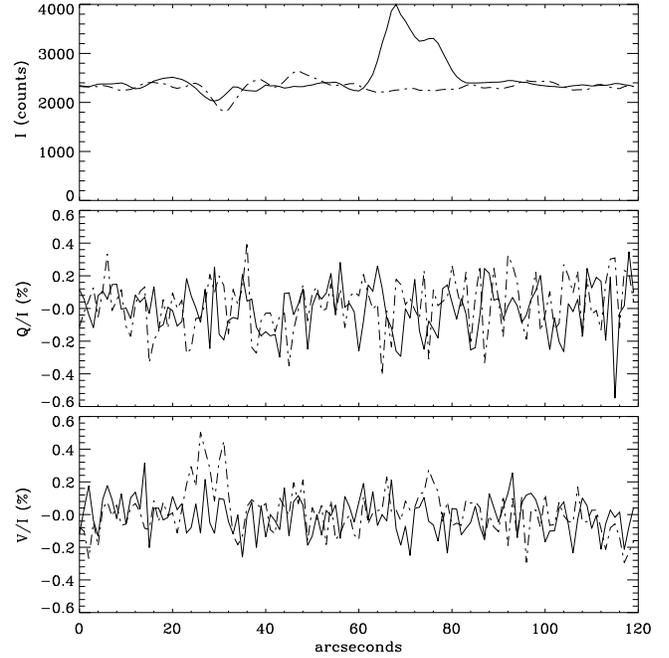}}
\end{center}
\vskip 0 cm
  \caption[]{Single spatial profiles taken from the 6th set of images of
Figure \ref{fig:1}. Horizontal rows are displayed passing through the
flare region (solid line) and through the sunspot (dot-dashed line).}
  \label{fig:2}
\end{figure}

Spatial Stokes profiles corresponding to a single row in the 6th set
of images of Fig.~\ref{fig:1} (bottom left) are shown in Fig.~\ref{fig:2}. 
The solid curves represent the profiles passing through the
flare region, 44\arcsec\ from the bottom of the image, while the dot-dashed
curves correspond to profiles through the sunspot, 18\arcsec\ from the bottom.
The noise level (rms) of single $Q/I$ exposures (2 seconds) is around
0.4\,\%, similar to all other flare observations. The over 10
exposures averaged polarization profiles reported
in Fig.~\ref{fig:2} allow to detect a $V/I$ signature.
The 0.3\,\% signal of the $V/I$ dot-dashed profile is a Zeeman
signature of the sunspot, due to the fact that the filter is
not perfectly centered in the H$\alpha$ line.
Thus any signature that exceeds a few tenths of a percent is visually
distinguished easily on the polarization images of Fig.~\ref{fig:1}.

For all filter mode observation data a visual examination of the time
evolution of the intensity and $Q/I$ frames is performed.  
Only in few flares we see small signatures. These are quite well
related to intensity gradients and for this reason we suspect an
instrumental origin.
Strong signatures, clearly exceeding several 0.1\,\%, were never
detected, but to give a measure of this results we proceed as
follows. We choose for each event a $10\arcsec \times 10\arcsec$ area
placed either where the intensity maximal value is detected or where
we visually see suspect $Q/I$ signatures.
The time evolution of the $Q/I$ values averaged over this area is
analyzed. 
The rms is calculated and given for each event in Table
\ref{listevents} as noise level. We consider also the running average
over ten exposures of $Q/I$ averaged over the small $10\arcsec \times
10\arcsec$ area, corresponding to an interval of 40 seconds. The
largest absolute value is reported in Table \ref{listevents} as
maximal polarization value. 
Where this value is missing in Table \ref{listevents} the measured
maximal polarization signature is less than the reported noise level.

\begin{figure}
\begin{center}
\leavevmode
\resizebox{8.5cm}{!}{\hspace{0cm}\vspace{0cm}\includegraphics{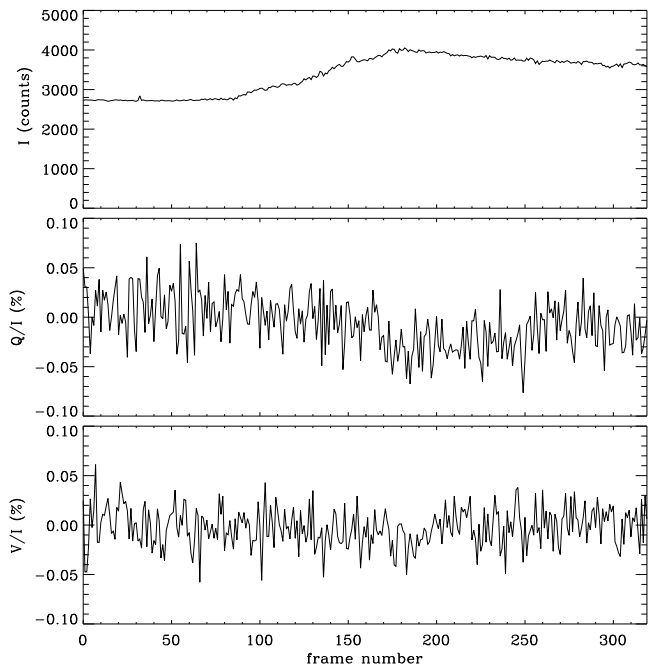}}
\end{center}
\vskip 0 cm
  \caption[]{The time evolution of intensity, $Q/I$ and $V/I$ averaged
over a small, 20\arcsec$\times$14\arcsec area in the flaring region
shown in Fig.~\ref{fig:1}.}
  \label{fig:3}
\end{figure}

As example we describe the reduction of the 7th July 2002 flare. The
$10\arcsec \times 10\arcsec$ area is centered at 65\arcsec\ from the
left and 40\arcsec\ from the bottom of the images reported in
Fig.~\ref{fig:1}. For each of the 320 frames we have the averaged
values over the small area for intensity, linear and circular
polarization, as reported in Fig.~\ref{fig:3}. 
The $Q/I$ evolution is almost unaffected by the flare event.

\begin{figure}
\begin{center}
\leavevmode
\resizebox{8.5cm}{!}{\hspace{0cm}\vspace{0cm}\includegraphics{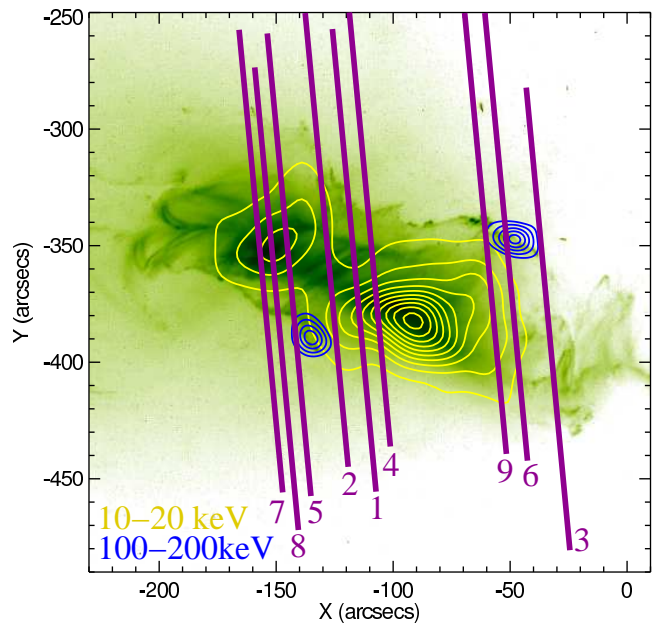}}
\end{center}
\vskip 0 cm
  \caption[]{Positions of the spectrograph slit on the intensity maps
  recorded by TRACE (195\,\AA, gray scale) and RHESSI (contour lines)
  for the region of the X17.1 flare of 28th October 2003 taken at
  11:11:53 UT. Soft X-ray contours are white, hard X-ray contours are black.}
  \label{fig:4}
\end{figure}

Also in the spectrograph measurements no significant signature could
be detected, which could be interpreted as impact polarization. As an
example we present an observation performed during one of the largest
flares ever observed: the X17.1 flare of 28th October 2003. Figure
\ref{fig:4} is a composite image of RHESSI and TRACE data, on which the
positions of the slit in the different observations performed with
ZIMPOL are drawn. 
The time intervals during which the event was observed at each
position are given in Table \ref{observ28oct}. The soft X-ray phase,
during which linear polarization has been claimed to be observed in
other flares, lasted from 10:32 UT until beyond 13:00 UT. All of the
slit positions except number 
3 are traversing the soft X-ray source.
The gamma ray emission with energy 0.8--7\,MeV was enhanced from 10:34
until the end of the RHESSI observing time at 11:25 UT.
Initial imaging results of the 2.2 MeV line emission integrated over
11:06-11:25 UT show two sources with a similar separation as seen in
Fig. \ref{fig:4} for hard X-rays (Hurford et al. 2004). 
The centroid positions of the 2.2 MeV line are slightly shifted ($\sim
15$\arcsec\ East, i.e. left in Fig. 4) relative to the hard X-ray centroids.

\begin{table}[h]
\caption[]{Time intervals of the observations at the different slit
  positions shown in Fig. 4.}
\label{observ28oct}
\begin{center}
\begin{tabular}{ccc} 
\hline \hline\noalign{\vskip2pt} 
slit position&    begin& end\\
 \noalign{\vskip2pt}\hline  \noalign{\vskip2pt}
1 &  11:10 &  11:20 \\
2 &  11:22 &  11:25 \\
3 &  11:26 &  11:29 \\
4 &  11:31 &  11:33 \\
5 &  11:34 &  11:36 \\
6 &  11:37 &  11:39 \\
7 &  11:42 &  11:44 \\
8 &  12:02 &  12:03 \\
9 &  12:07 &  12:08 \\
\noalign{\vskip2pt}\hline
\end{tabular}
\end{center}
\end{table}                 

\begin{figure}[ht]
   \centering
   \includegraphics[angle=90,width=10cm]{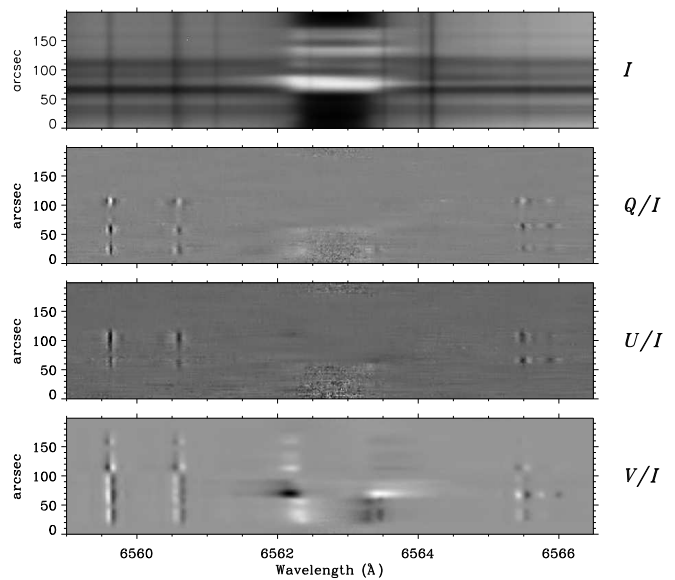}
      \caption{Full Stokes spectral images taken with the spectrograph
      slit in position 5 of Fig.~\ref{fig:4}.}
         \label{fig:5}
   \end{figure}

\begin{figure}
\begin{center}
\leavevmode
\resizebox{8.5cm}{!}{\hspace{0cm}\vspace{0cm}\includegraphics{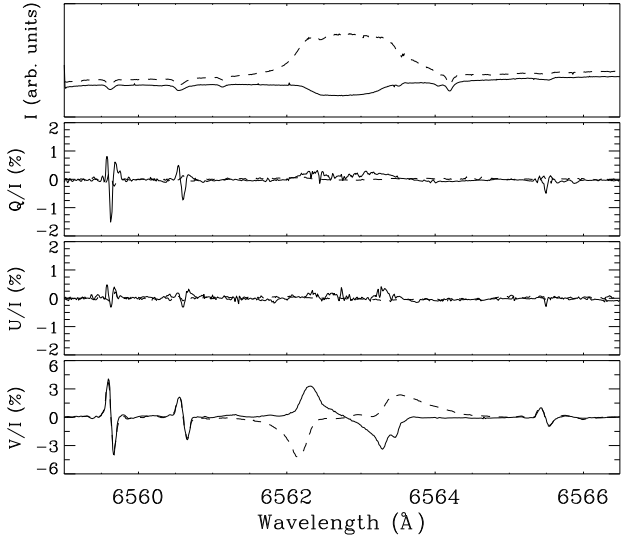}}
\end{center}
\vskip 0 cm
  \caption[]{Spectral Stokes profiles taken from two different
  horizontal rows of Fig. \ref{fig:5}. A position with strong flare
  emission is shown dashed, while a quiet position is represented by
  the solid curves.}
  \label{fig:6}
\end{figure}

Figure \ref{fig:5} shows the result of a 10\,s measurement
obtained at slit position 5 at 11:34 UT.  
The intensity image is corrected for flat field. 
The $Q/I$, $U/I$ and $V/I$ images are smoothed with the wavelet technique
(Fligge \&\ Solanki 1997). 
In the linear polarization images we find the usual symmetric Zeeman
signatures in the atomic solar lines but not in the telluric
lines. They can be seen as an expression of the polarimetric
accuracy. In the H$\alpha$ line however, no signature is detectable
that could indicate impact polarization. An increase of the noise due
to low intensity can be noticed in the lower part of the H$\alpha$
line. 
Correlated with the strong intensity gradient in the bottom part 
of the flare region, a small signal below 0.3\,\%, probably of 
spurious origin, can be noticed (cf. Fig.~\ref{fig:6}). 
In the $V/I$ image the H$\alpha$ circular polarization changes 
sign in the flare emission region, in contrast to the surrounding 
absorption lines.

Figure \ref{fig:6} shows spectral profiles extracted from
Fig.~\ref{fig:5} for two different spatial pixels. The solid lines
correspond to the horizontal row at 59\arcsec\ from the bottom,
where the largest linear polarization
signal in H$\alpha$ is detected. It includes
the region with the largest intensity
gradient.
The dotted line refers to the position with 
maximum H$\alpha$ intensity (horizontal row at 77\arcsec). 

All other observations taken during this and the other measured flares
give the same null results for impact polarization.
Signals similar to those seen in the linear polarization in
Fig.~\ref{fig:6}, which could be related to spurious signals as
mentioned in Section 2.3, are detected in few cases in other
spectrograph data. 
This signatures are correlated with the intensity gradient and thus
have negative and positive signs that partially cancel when averaging.
A rebinning of the spectral $Q/I$ images, averaging over $4 \times 4$ pixel,
cancels this spurious signatures. The rms of the $Q/I$ spectra is
reported in Table \ref{listevents} as noise level. These values are
calculated in a 10 seconds exposure $Q/I$ measurements, the slit of
the spectrograph was subtending 1.5\arcsec\ and a pixel element (after
rebinning) has a size of $5\arcsec \times 40$ m\AA.
Large rms values are found when observing outside the disk.

\section{Discussion}
No linear polarization in H$\alpha$ has been found in our
measurements of 30 solar flares from GOES class C2.1 to X17.1. These 
events were associated with hard X-ray emission when RHESSI was 
observing, one of them even in gamma-ray lines. Non-thermal electrons 
and possibly protons must have been accelerated in these flares, 
propagated to the chromosphere and impacted on hydrogen atoms. 
The question thus arises why impact polarization was not
detected. 

Impact polarization has been observed in the laboratory and agrees
with calculations from first principles (Percival \& Seaton 1958). 
For a review, the reader is
referred to Vogt \& H\'enoux (1996). The degree of polarization $P$
depends on the viewing angle $\beta$ between the impacting particle's
velocity and the observer, and on the energy $E$ of the particle: 
\begin{equation}
P(\beta,E) = b P_{90}(E)\, {\sin^2\beta\over{1-P_{90}(E)\cos^2\beta}}\ \ \ ,
\end{equation}
where $P_{90}(E)$ is the polarization at $\beta=90^\circ$ for a 
fully collimated beam. The polarization peaks at a perpendicular 
viewing angle and vanishes for a
parallel line of sight relative to the particle motion. Following 
Fletcher \& Brown (1995) we take $P_{90}=0.246$ of 1 MeV  protons 
(3p-2s transition). This value must be multiplied by the anisotropy 
factor $b<$1, ranging from 0 (fully isotropic), about 0.5 for a 
cosine-distribution in pitch angle, to 1 for a unidirectional 
beam (H\'enoux et al. 1983). 

The interpretation of the previously reported alleged
detections of linear polarization favored proton beams as the cause. 
This is consistent with reported evidence for flare accelerated
protons. Nuclear excitation lines in gamma-rays are produced by high
energy protons impinging on the chromosphere (Ramaty \& Murphy
1987). In one of the flares reported here (Fig.~\ref{fig:4}), RHESSI
has indeed recorded clear signatures of protons. Theoretical models
based on proton fluxes consistent with such observations have been
published by Vogt et al. (1997). They predict linear polarizations of
a few percent, assuming a fully collimated proton beam.

The absence of an observable degree of linear polarization may have
two reasons: {\it (i)} The energetic protons may not reach the
atmospheric level of the H$\alpha$ emission in sufficient numbers, 
or {\it (ii)} the protons may be nearly isotropic when they 
reach that level. 

The first possibility may be excluded in view of the moustache
phenomenon. In certain large flares there is sometimes a phase when
the H$\alpha$ line is broader than 10\,\AA. This suggests that
individual hydrogen atoms move with velocities of more than $10^7$\,km
s$^{-1}$, as one may expect after collisions with energetic particles
(Ding, H\'enoux, \& Fang 1998). We have therefore examined with
special care the linear polarization for such events. One such case
was presented in Sect.~\ref{sec:results}. The circumstance that no
polarization above threshold was found suggests that the exciting
particles (if present) were not beamed. This is consistent with the
fact that the H$\alpha$ line profile is not broadened in only the red
direction during the moustache phase.

In the models quoted above, the protons are assumed to be accelerated
isotropically in the corona. A beam develops as protons propagating in
the plasma are slowed down by electron friction (H\'enoux et
al. 1990). The loss of energy is proportional to the distance
traveled. Protons propagating parallel to the guiding magnetic field
travel the shortest distance, while the others follow a path that is a
factor of $1/\cos\alpha$ longer, where $\alpha$ is the pitch angle
between the proton velocity and the magnetic field. Obviously, the
resulting beam is not strongly collimated, contrary to some models
assumptions. Collimation is further limited by the beam instability to
growing Alfv\'en waves. If the instability occurs, the protons 
become nearly isotropic and propagate parallel to the magnetic 
field with a mean velocity of the Alfv\'en speed (Wentzel 1974).

In addition, there are processes that work in the opposite direction,
trying to randomize the angular distribution. For instance,
interactions with neutrals scatter the protons efficiently. The 
scattering time $\tau$ of a proton with energy of 1 - 500 MeV in 
a weakly ionized gas is
\begin{equation}
\tau = - {E_{\rm MeV}\over{dE_{\rm MeV}\over dt}} \approx 
{E_{\rm MeV}^2\over 3\cdot 10^{-12} n_H \beta 
(\ln E_{\rm MeV} + 22.3)}\ \ ,
\end{equation}
where $E_{\rm MeV}$ is the kinetic proton energy in MeV and 
$n_H$ is the hydrogen density in cm$^{-3}$ (Benz \& Gold 1971, 
after Hayakawa \& Kitao 1956). With a 
$n_H \approx 6\cdot 10^{13}$cm$^{-3}$ in the H$\alpha$ 
emitting region, the mean free path for a proton of 1 MeV 
for example becomes 75 km.  Thus
collisional focusing is limited to the corona and the transition
region, and works in the opposite sense in the
chromosphere. 

Isotropization is enhanced by a second effect based on
conservation of the magnetic moment. Charged particles are forced into
more transverse orbits as they propagate downward along a converging
magnetic field. The pitch angle $\alpha$ is related to the local
magnetic field $B$ by

\begin{equation}
\alpha = {\rm arcsin}\left( \sin \alpha_0 
{\sqrt {B\over B_0}}\right) \ \ \ ,
\end{equation}
where index 0 refers to a site of reference along the orbit 
or the initial value at the release from the acceleration region. 
The effect widens up a beam that has an initial pitch angle width 
of $\alpha_0^{\rm max}$ to $\alpha^{\rm max}$ at a place with 
magnetic field $B > B_0$ according to Eq.(3). Let the initial 
velocity distribution be
\begin{equation}
f_0(v,\alpha_0)\ =\ f(v)H(\alpha^{\rm max}_0 - \alpha_0) \ \ \ ,
\end{equation}
where $v$ is the absolute value of the velocity (which is 
conserved) and $H$ the Heaviside function. At the local 
magnetic field $B>B_0$ the pitch angle $\alpha$ is given by 
Eq.(3), and
\begin{equation}
f(v,\alpha)\ =\ f(v)H(\alpha^{\rm max} - \alpha){\cos\alpha 
\over \sqrt{(B/B_0-1)+\cos^2\alpha}} \ .
\end{equation}
Thus an increase by a factor of 4 in the magnetic field 
strength from the acceleration site to the H$\alpha$ emitting 
region would enhance a
small pitch angle by approximately a factor of 2 (Eq.3). The proton beam
widens up in velocity space by the same factor. Scattering by neutrals
and conservation of the magnetic moment may act together to reduce the
anisotropy to a level where impact polarization becomes negligible.


Other causes for linear polarization by impact polarization at 
the one percent level have been proposed, such as by the downgoing 
heat flux (H\'enoux et al. 1983), by evaporative upflows 
(Flechter \& Brown 1998), and by return currents (H\'enoux and 
Karlicky 2003). They apparently also overestimated the effect 
trying to match the previously reported values.

\section{Conclusions}
Linear polarization values up to 20\% and more have been reported 
in the literature for more than 20 years. We have not been able 
to confirm these values with an imaging system that recognizes 
flares, registers both the preflare and flare phases at noise 
levels of some 0.1\%. The fast polarization modulation and special 
instrumental precautions  
much reduce the level of spurious effects from seeing and instrumental 
causes compared to all previous observations. 

Our data set includes 30 flares of great variety, varying from small 
to very large by two orders of magnitude in peak soft X-ray flux. 
They occurred at different positions on the disk, some near the 
limb or even beyond the disk. The observations were made with 
high time resolution and cover all the flare phases. We have used 
both imaging and spectrographic polarimetry.

None of flares showed more than 0.07\% linear H$\alpha$ polarization 
in a $10\arcsec \times 10\arcsec$ area during an integration time of 
40 seconds. The above peak value and all others are not significantly 
different from zero.

These observations do not exclude impact polarization in general, 
but contradict all previous positive reports. Our theoretical 
considerations demonstrate that the anisotropy of energetic
ions at the site of H$\alpha$ emission has been largely overestimated 
in the past to interpret the previously reported polarization levels. 
Impact polarization may exist only if it is limited to a smaller area 
or shorter time. Future measurements of solar flare linear 
polarization may still be meaningful if done at the present noise 
level of some 0.1\%, excluding seeing effects and instrumental 
polarization, and with enhanced spatial and temporal resolution.

\begin{acknowledgements} 
We are grateful for the financial support that has been provided by
the canton of Ticino, the city of Locarno, ETH Zurich, the Swiss
Nationalfonds, the Hessisches Ministerium f\"ur Wissenschaft und Kunst
(HMWK), and the foundation Carlo e Albina Cavargna. Peter Steiner and
the ZIMPOL team helped to update the ZIMPOL software for these
measurements. S\"am Krucker has provided the RHESSI and TRACE images
for Fig.4. 
Reiner Klein, University of Applied Sciences (Fachhochschule) in Wiesbaden,
developed the software of the automated flare detection system.
\end{acknowledgements}



\begin{thebibliography}{}
\bibitem{} Benz, A.O., Gold, T. 1971, Solar Phys., 21, 157 
\bibitem{} Bianda, M., Stenflo, J.O., Gandorfer, A. M., Gisler, D.,
  K\"uveler, G. 2003, in: Solar Polarization 3, ASP Conf. Ser. 307, eds. J. Trujillo Bueno, S\'anchez Almeida, p. 487
\bibitem{} Canfield, R. C., Chang, C.-R. 1985, ApJ, 295, 275
\bibitem{} Ding, M. D., H\'enoux, J.-C., Fang, C. 1998, A\&A, 332, 761
\bibitem{} Fang, C., Feautrier, N., H\'enoux, J.-C. 1995, A\&A, 297, 854 
\bibitem{} Firstova, N.M.,\& Boulatov, A.V. 1996, Solar Phys., 164, 361 
\bibitem{} Fletcher, L. \& Brown, J.C. 1995, A\&A, 294, 260
\bibitem{} Fletcher, L. \& Brown, J.C. 1998, A\&A, 338, 737
\bibitem{} Fligge M.,\& Solanki  S.K. 1997, Astron. Astrophys. Suppl. Ser., 124, 579  
\bibitem{} Gandorfer, A. M., Povel, H. P., Aebersold, F., Egger, U., Feller, A., Gisler, D., Hagenbuch, S., Steiner, P., 
Stenflo, J. O. 2004, A\&A, 422, 703 
\bibitem{} Hayakawa, P., Kitao, K. 1956, Prog. Theor. Phys., 16, 139
\bibitem{} Hanakoka, Y. 2003, ApJ, 596, 1347
\bibitem{} H\'enoux, J.-C., Chambe, G. 1990, J. Quant. Spectrosc. Radiat. Transfer, 44, 193
\bibitem{} H\'enoux, J.-C., Chambe, G., Tamres, D., Feautrier, N., Rovira, M., Sahal-Br\'echot, S. 1990, ApJS, 73, 303
\bibitem{} H\'enoux, J.-C., Karlick\'y, M. 2003, A\&A, 407, 1103
\bibitem{} Hurford, G.~J., Krucker, S., Lin, R.~P., Schwartz, R.~A., Share, G.~H., Smith, D.~M. 2004, AAS meeting, 204, 2.02
\bibitem{} Kashapova, L.K. 2003, in: Solar Polarization 3, ASP Conf. Ser. 307, eds. J. Trujillo Bueno, S\`anchez Almeida, p. 474
\bibitem{} K\"uveler G., Klein, R., \& Bianda, M. 2003, Photonik, 2/2003, 66
\bibitem{} Metcalf, T., Wulser, J.-P., Canfield, R. C., Hudson,
  H. 1992, in: The Comption Observatory Science Workshop, Vol. NASA
  Conf. Proceedings, 3137, 536
\bibitem{} Metcalf, T., Mickey, D., Canfield, R., Wulser, J.-P. 1994, in : High Energy Solar Phenomena, Vol. AIP Conf. Proceedings, 294, 59
\bibitem{} Percival, I.C., Seaton, M.J. 1958, Phil. Trans. Roy. Soc. London, Series A, 251, 113
\bibitem{} Povel, H. 1995, Optical Engineering, 34, 1870
\bibitem{} Ramaty, R., Murphy, R. J. 1987, Space Sci. Rev., 45, 213
\bibitem{} S\'anchez Almeida J., Martinez Pillet, V., \& Wittmann, A.D. 1991, Solar Phys., 164, 361
\bibitem{} Wentzel, D.G. 1974, Ann. Rev. Astr. Ap., 12, 71
\bibitem{} Vogt, E., H\'enoux, J.-C. 1996, Solar Phys., 164, 345 
\bibitem{} Vogt, E., Sahal-Br\'echot, S., H\'enoux, J.-C. 1997, A\&A, 324, 1211
\bibitem{} Vogt, E., H\'enoux, J.-C. 1999, A\&A, 349, 283 
\bibitem{} Vogt, E., Sahal-Br\'echot, S., H\'enoux, J.-C. 2002, in:
  Proceedings Second Solar Cycle and Space Weather Euroconf., ESA
  SP-477, 191 
\bibitem{} Wiehr, E. 1971, Solar Physics, 18, 226

\end{thebibliography}
\end{document}